# TOTAL NOVEL AND COMPLEXITY.
# Literature and Complexity Science

Carlos Eduardo Maldonado
Full Professor
School of Medicine
Universidad El Bosque
ORCID: http://orcid.org/0000-0002-9262-8879
maldonadocarlos@unbosque.edu.co

Abstract

A strong link between complexity theory and literature is possible, i.e. feasible, under one proviso, namely that total novels be considered. However, neither in literature at large nor in complexity science has been literature seriously taken into consideration. This paper argues that a total novel is most conspicuous example of a complex system. The argument is supported by a clear characterization of what a total novel is and entails. Science and literature can be thus complemented and developed, hand in hand.

Introduction

The sciences of complexity have largely contributed to the understanding of our world, reality, and the universe when taken as complex systems or when exhibiting complex behaviors. A number of properties have been highlighted that help characterize what a complex system is, such as nonlinearity, emergence, turbulence, the existence of power laws, fluctuations, self-organization, and many others.

In the rare occasions where complexity theory and the humanities and arts at large have been brought together, the emphasis has been placed on the representational arts (Casti and Karlqvist, 2003). A careful search on the relations between complexity an literature brings out two studies (Zants, 1996; Hess, 1999). Surprisingly, these two books had little impact both in the academic, scientific and humanist communities. Albeit they link literature and complexity science, there is not one single word about total novels. That is why they must remain here out of scope.

It is the contention of this paper that there are nearly no studies relating complexity and literature. This paper aims at filling in such a gap. In order to do so, the claim will be made that the best link between complexity theory and literature arises when considering a total novel. The argument then will be developed throughout three sections, thus: firstly, provided that there are nearly no explicit elements, neither in literature theory nor in literature critique, about the character of a total novel some elements for a theory of a total novel will be brought. Accordingly, some elements for a theory of a total novel will be sketched here. The first section alone justifies the novelty of this paper. Secondly, the justification of a total novel as a complex system will be highlighted. A total novel can be seen as sort of epitome in literature. A simulation of a total novel is introduced herein. The third section focuses on the relationship between science, i.e. complexity science and literature. It is a striving for a generalization that shed some new lights on the very understanding of complexity theory. Finally, some conclusions are drawn.

1-. Some elements for a theory of a total novel

In literature a new genre has arisen recently. The framework of its appearance was the so-called Latin American "Boom" in the 1960s and 1970s (Forero Quintero, 2011; Garcés Valenzuela, 2010). Such genre has been coined as "total novel".

To date, a list of total novels undoubtedly includes the following novels (alphabetically considered by author) (See Olmor, 2014):

- Roberto Bolaño, *2666*
- Julio Cortazar, *Hopscotch*
- Miguel de Cervantes, *Don Quixote*
- Fyodor Dostoievsky, *The Karamazov Brothers*
- Carlos Fuentes, *Terra Nostra*
- Gabriel García Márquez, *A Hundred Years of Solitude*
- Pierre Jourde, *The Absolute Maréchal*
- James Joyce, *Ulysses*
- Joanot Martorell, *Tirant, The White*
- Lev Tolstoi, *War and Peace*
- Mario Vargas Llosa, *The Green House*

The list is to be enlarged also by:

- Fernando del Paso's, *José Trigo*, *Palinuro of Mexico*, *News from the Empire*

A number of characteristics of a total novel pump up immediately to a sensitive glance, thus: each total novel is a large and dense work (if taken by the number of pages, f.i.). It has a central thread that nonetheless bifurcates –literally, exactly in the sense of the bifurcations studied by Prigogine or Wolfram, for example. Numerous situations, characters, stories and timelines move along the central thread, complement it, diverge, are developed in parallel, and many times they do not have much to do with the central character of the novel, and yet help understand the whole story.

There is, albeit, not a fully developed theory of a "total novel" - neither from literature theory nor from literature critique or criticism. Rather, its characterization is a matter of taste, descriptions, and inner experience, very much like a phenomenological stance, an approach very much used in studying the sciences of complexity.

Being as it may be, a total novel is a multi-layer story, very much like a geological structure, or the rings of a tree, i. e. each layer tells a different story, has a different temporality, and nonetheless all the different layers make up a whole that is never entirely achieved – for it is a living phenomenon. In one word, a total novel is a multi-scale system.

A total novel is an aesthetic maze of stories, pivoting, so to speak around a central one. Sometimes time elongates as if on its own, and sometimes time is a variation on a central theme, very much like Bach's variations. It is just a matter of case, and author to locate, confirm and enjoy these characteristics. To be sure, a total novel is not a linear story – in case there are linear novels, at all – but explicitly a nonlinear set of a number of stories, each one overlapped on the others, each crossing other(s), having some bifurcations that lead nowhere (just like in life, in fact), some fully depicted while others are just sketched or barely suggested.

A good counter-examples help illustrate what has been said, so far. M. Proust's *In Search of Lost Time* one the one side, but also Lady M. Shikibu's marvelous novel *The Tale of Genji* are not considered as total novels, in spite of their length and complicated structure, and beauty. Several other examples could be mentioned along the same tenure, past or current.

A total novel is the outcome of a subtle and well-made dynamic balance between fiction and essay, prose and poetry, logos and myth, imagination, and anthropology and history. It is neither more fiction nor sharp analysis of reality, and not more fantasy than research and documentation. Instead, a total novel is an intelligent and creative mixture of both components, the work of a thinker as well as of an artist.

In an interpretation, literature should be more than literature and be able of a real contribution to the understanding of the world, reality, and nature (Ruiz-Pérez, 2016).

In a total novel the borderline between normality and fantasy (if not madness) is blurry, fuzzy, mobile. *Don Quixote* is, by and large, the best example. "Magic realism", was the concept coined to describe the thinking and writing style by authors such as García Márquez and Vargas Llosa, among others. The linearity of time becomes questionable, almost non-existent. This is exactly what happens to Leopold Bloom in Joyce's *Ulysses*, for instance, not to mention the avatars of time in F. del Paso's *Palinuro of Mexico* (Saenz, 1994; Anderson 2015). The use of digressions, flashbacks, shifting time, and the displacement of space are some of the generic constitutive traits a total novel bears.

As it can be clearly seen, thanks to its very nature, a theory of a total novel does comprise a complexity approach, from one extreme to the other, so to speak. All in all, a total novel is a most fascinating complex system – something that has not been fully grasped neither on the side of complexologists, nor on the side of literature theory and critique.

In any case, it should be clear, that there is not a canon to a total novel. Moreover, each total novel is singular and unique. That is exactly its greatness. If so, a theory of a total novel is not a universal formal and abstract theory, but a theory of a particular phenomenon – in each case. Very much, alas, like a theory of living beings (can we remind that to date there is nothing as a universal theory of life or living beings?).

In one word, a total novel is a complex worldview (*Weltachanschaung*), This means, that either an entire comprehension of the world as it happens can be achieved (García Márquez's *A Hundred Years of Solitude* or C. Fuentes' *Terra Nostra*) (Anderson, 2003), or also that a whole understanding of the human heart and mind has been gotten (Dostoievsky's *The Karamazov Brothers*).

A theory of a total novel sets out the general traits of a wonderful and singular novel that is irreplaceable, by definition. Yet, a modeling or simulation of a total novel is possible and desirable. Here is when we turn on to the next section.

2-. Complexity and Total Novel

In the following, based on NetLogo®, a simulation of the origin and development of a total novel is presented. In the first frame, a story originates that has bifurcations, with several situations, characters, and places or times, that are the consequence of an original setting. Of course, it is not necessary that the story be understood or explained from the onset, because it can be grasped as it unfolds, as it happens.

**Frame No.1 of Simulation: A Story Begins and Unfolds**

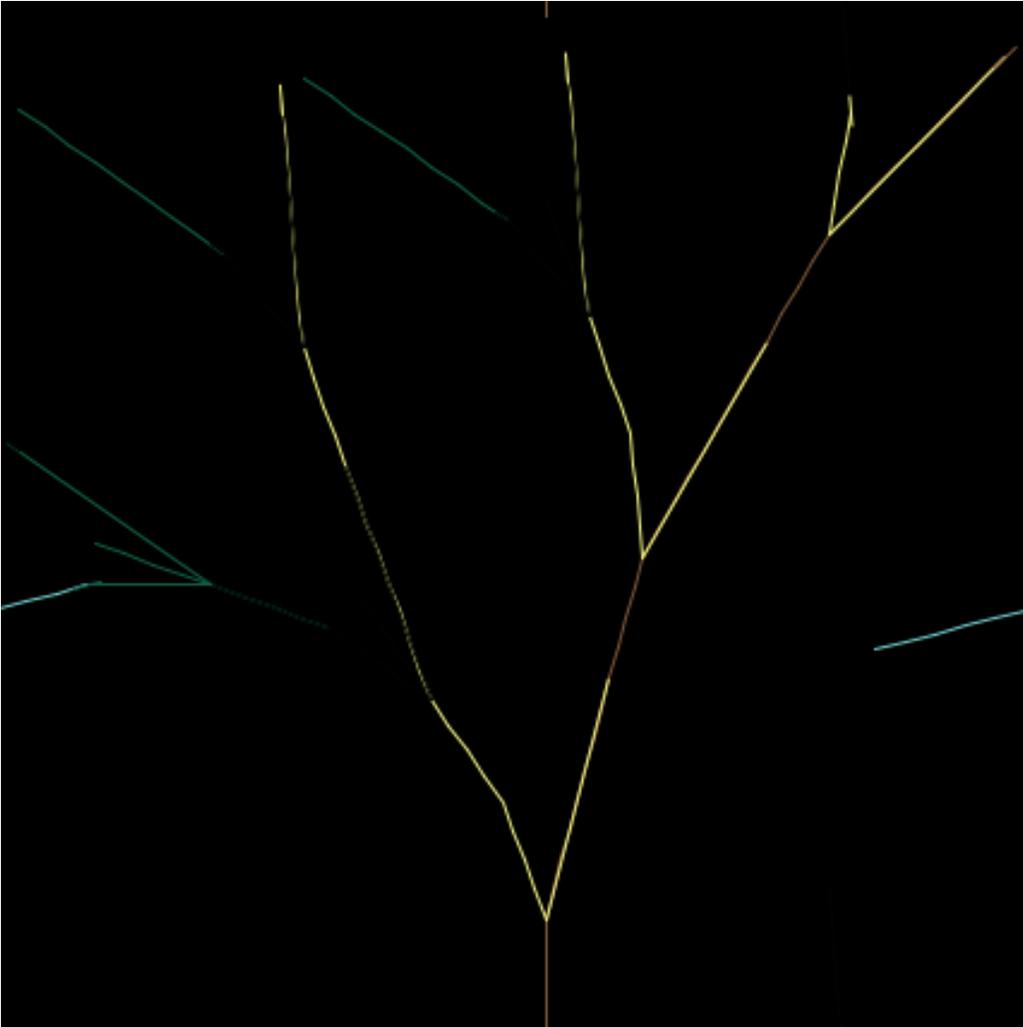

Source: Own elaboration, after a model by Wilensky (1997) and NetLogo®

As it can be seen, there are unforeseen stories, situations, events or timelines that emerge from the central tree. Those emergent phenomena can fulfill, bring sudden developments, or enlarge the scope of a given circumstance. In frame No. 1 the unforeseen stories and events are depicted in blue, and arise from the margins and are going to cross the central story developed so far.

In the language of NetLogo®, 8 turtles are present in frame 1. The different buttons of the simulations are omitted here that help model the story (sliders, etc.).

**Frame No. 2 of Simulation: A Complexification of the Story Takes Place**

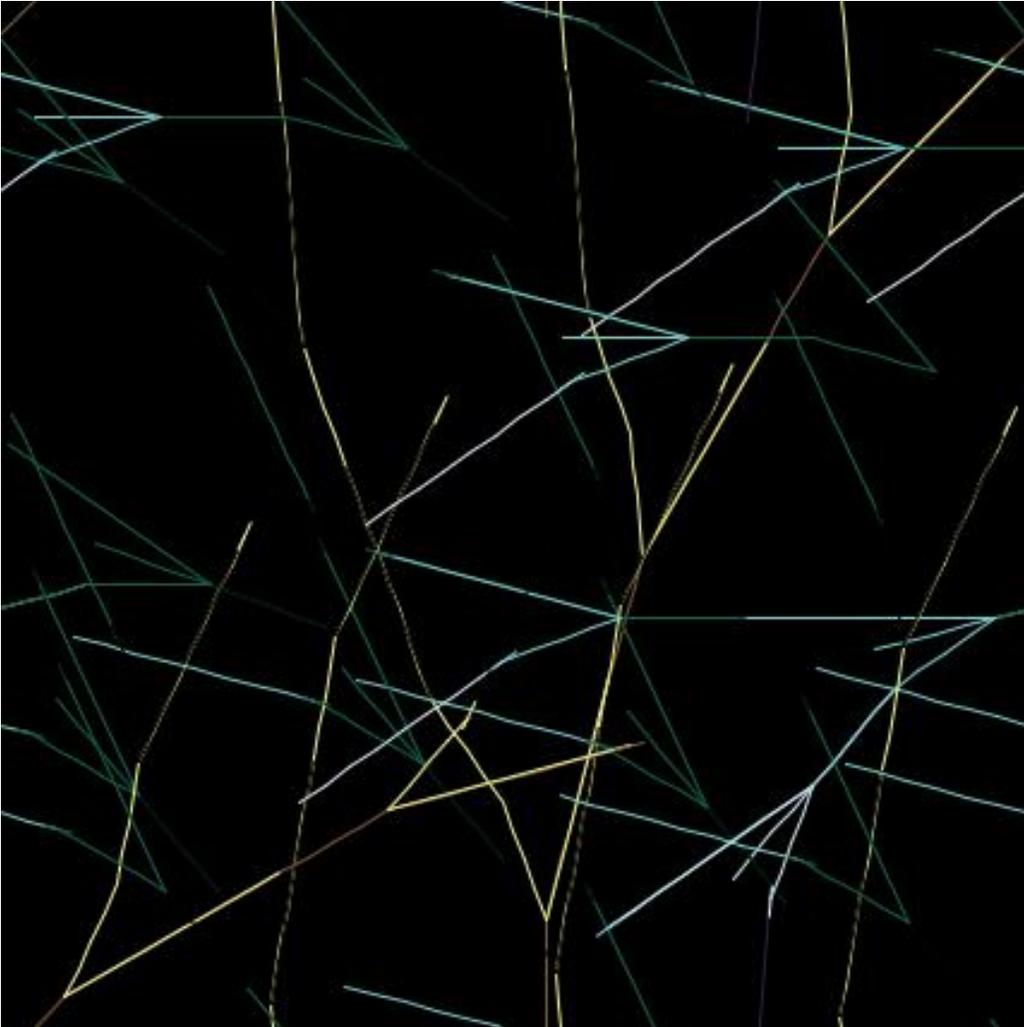

Source: Own elaboration, after a model by Wilensky (1997) and NetLogo®

In Frame No. 2, a series of fortuitous developments take place that make the story always more complex. Drama and comedy are intertwined; tragedy and absurd are subtly mixed in such a way that there is not a clear borderline between them. Numerous situations, events, characters and displacements take place that render the novel ever more complex. Sometimes, it is not clear what exactly an event means in relation with the main core; yet, such an event is self-contained. The event can be a reflection on society, on history, on politics, on the universe or on the human heart. The author is little by little reckoned as a psychologist as well as a creator of fictional characters. Nonetheless, the border between fiction and reality becomes permeable, and moves constantly here and there.

In the language of NetLogo®, 32 turtles are simulated in frame No. 2. Here too, the various buttons of the program are left aside.

**Frame No. 3 of Simulation: A Total Novel is Accomplished**

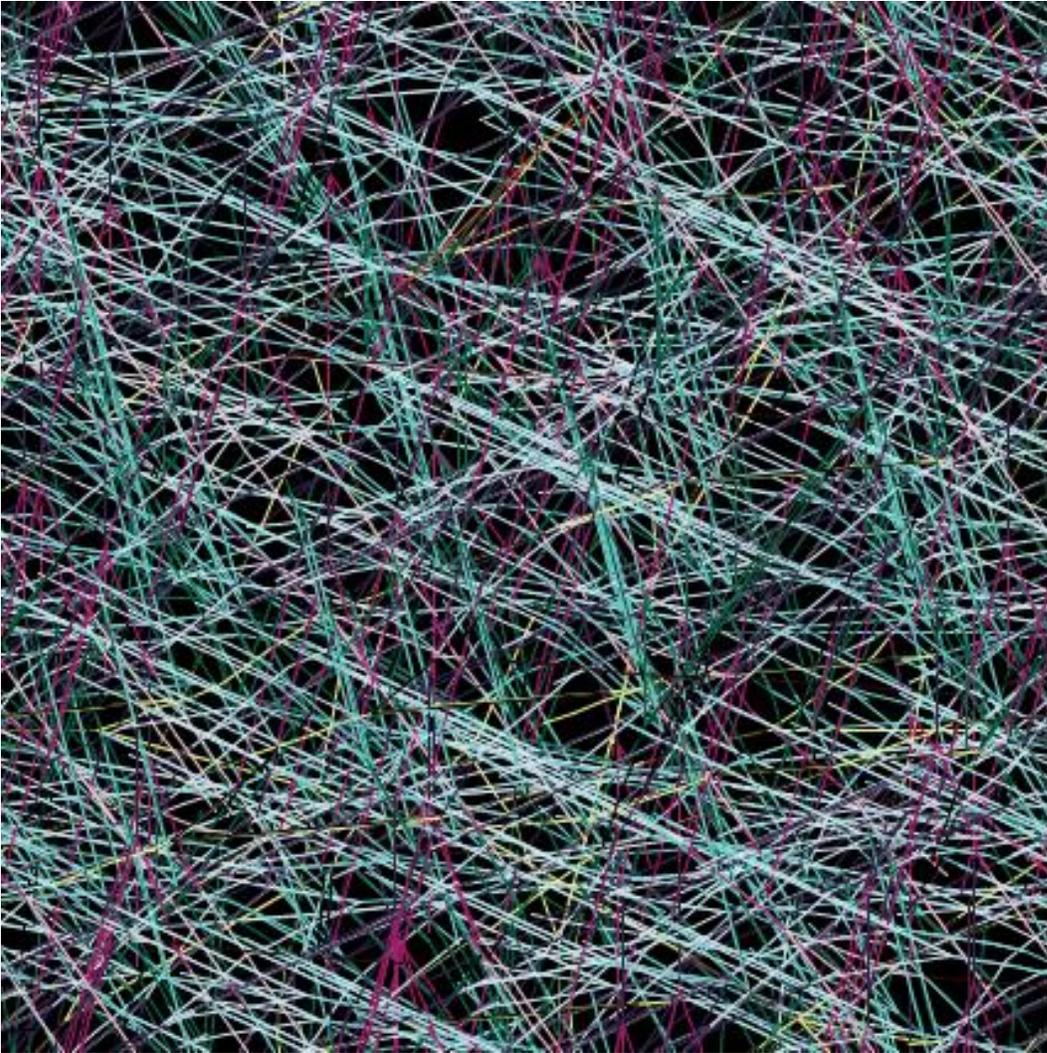
Source: Own elaboration, after a model by Wilensky (1997) and NetLogo®

Frame No. 3 shows a complex panorama, a real maze of many stories, events, situations, and characters that make up a whole deep story. The original development seems to be forgotten, and the early situations seem to be left far behind. Yet, they help configure a most fantastic and dense work that the reader can enjoy and in which, literally, the reader has been trapped. Such is the ingenuity of the writer, and the author can be acknowledged as a great author. A total novel has emerged that produces delight and passion that moves the reader and shocks standards and common places.

As it can be seen, a total novel permits a set of gaps, ambiguities, ambivalences, and situations taken for granted. These are depicted in the black spots that remain unfulfilled in the back of the screen. The idea here is that a total novel is not a totalizing work, for instance in the sense of Hegel's *Phenomenology of the Mind*, or on a quite different context, S. Wolfram's *A New Kind of Science*. A total novel is, indeed, a self-contained universe, yet leaving some implicit knowledge and developments here and there, that the reader must be able to imagine and fulfill. Literature in this sense exhibits some freedom that seems to be more reduced or constrained in science or philosophy.

The simulation presented in frame No. 3 is made up with 512 turtles, in the language of NetLogo®, and it aims at depicting a most complex work. It goes without saying that the number of turtles can be increased at will. Yet, for the purposes of this study, the number can be considered as satisfactory.

All in all, a total novel can be adequately said to be a complex system in that non-linear developments, bursts, and parallel and distributed layers come to complement each other in a way that was unexpected from the onset. Unpredictability, emergence, and free-scale networks appear as clear references that are to be specifically illustrated when considering each total novel, and author.

A total novel is not just a literary work; rather, it is an encompassing work with high caliber that enables the reader to smoothly move from reality to fantasy, from her own experiences to the world around and history, from laughing to sorrow, from deep concern to surprise and astonishment (Corral, 2001). Language, a most intelligent language, accompanied by good research and data, are gracefully intertwined in a picture that teaches as well as distracts, brings insight and produces relief and a strange, singular form. No one leaves a total novel untouched or in the same tenure as the reader had when first digging into it, even if, as it sometimes happens, the reader has to stop reading the novel for one reason or another.

3-. (Complexity) Science and Literature

According to the western tradition ever since Aristotle there are differences in writing and thinking styles. Such difference, though, did entail a sort of hierarchy, in the sense that there are better or more accomplished forms of knowledge and writings than others. From the point of view of a science and philosophy of complexity such Aristotelian claim becomes questionable, if not untenable.

The difference between science and the arts, in general and particularly between science and literature is much more subtle than it appears. Both science and literature do rely on data, i.e. observations. Much more meaningfully both are nurtured by imagination, if not also by insight and intuition.

From another standpoint, science both leads and relies on beauty (Tiezzi, 2000), as most notably mathematicians would be most willing so assess, whereas beauty is the atmosphere, so to speak, of words and meaningful wording.

Literature in general is culturally fundamental insofar as it is about good storytelling. From the cultural standpoint, what remains at the end of the day is no the equation, the demonstration, or the approval or refusal of a theory – but the story told. Any good scientist must be a good story teller, no matter what. Science is not made up of concepts, categories, proofs, and judgments. Besides, science is also made up with metaphors, similes, foreshadowing, analogies, hyperboles, oxymorons, alliterations, allusions, and many other *tropoi* that arise from the side of poetry, the arts, and literature. It is the intelligent mixture of concepts and metaphors, of proofs and epigraphs, for example, what captivates a sensitive mind, but also forms it.

Complexity science has produced a large set of innovations in knowledge, at large. The first face of those novelties is language, namely brand new concepts, terms, names, and metaphors that shed new lights on the panorama of the universe, reality, and the world. The list would be too large, but it includes power laws, free-scale networks, synchronicity, emergence, chaos, turbulence, catastrophes, bursting, swarm intelligence, self-organization, degrees of freedom, dissipative structures, attractors, first order and second order transitions, and many others.

Such new language introduced by complexity theory at large allows to see and to explain brand new systems, phenomena and behaviors never before considered or understood. Complexity science has pervaded biology and mathematics, chemistry and computation, sociology and economics, physics and cosmology, among other fields, sciences and disciplines. Yet, there seems to be a gap between science at large, and the humanities (social and human sciences). It would take a different paper to sketch the state-of.-the-art of the relationship between complexity science and the arts. To be sure, that would be a fruitful but short (review) paper.

The point here consists in bringing together two domains that have been traditionally apart, since Aristotle, namely, science and literature. Interdisciplinary or cross-disciplinary approaches are necessary when trying to understand complexity – a complex world in times of turbulence. Total novels are more than suitable complex systems that enable the very task and endeavor of complexity theory – under the proviso, that the hierarchies and distances between them be vanished or banned.

In one word, good science is very good storytelling. Let's just take the time necessary for the story to be well told.

4-. <u>Some Conclusions</u>

There is a danger when studying the sciences of complexity, namely the terrain might become sloppy and scholars and researchers might slip into sort of scientism; in other words, assuming that sciences are different to humanities, falling back into the two cultures dilemma.

Thinking in complexity does not entail ensuring the idea of hierarchies, but of connections, networks, interdependence, symbiosis, cooperation. If so, then science at large can learn from literature and the arts, and vice versa. From the standpoint of culture, scientists are to be good storytellers. However, the structure of science based on papers does not allow for good stories to be told, and certainly not total stories. Books are much more feasible to open up the gate for a good and total story to be told. Each complex system has its own complexity, and probably there is not a universal formal characteristic for all complex systems. This issue remains as one of the most sensitive and susceptible to discussions within the framework of the sciences of complexity. The last word has not been said, so far, but the first seeds for its boarding can be clearly located (Cowan, *et al.*, 1999).

There is nothing as the total novel, but an array, suggestive and provocative of total novels. If the deep inner dream of a scientist consists in reaching an *eureka* moment, the deep inner strive of a writer is to write a total novel. However, deep inner dreams must never be told loudly beforehand. Each thinker, researcher or writer assumes his or her deep inner dream as a long-lasting effort that can be reached in due time. When that happens, the *aha*! moment becomes one and the same thing with the life of the researcher, thinker or writer. Then a *work* is accomplished, and not just a paper or a book.

Bibliography

Anderson M. D., (2015). "Modernism, crisis, and the ethics of democratic representation in Fernando del Paso's total novels", in: *Latin American Research Review*, vol. 50, No. 2, pp. 42-62, The Latin American Studies Association

Anderson, M., (2003). "A reappraisal of the "total" novel: Totality and communicative systems in Carlos Fuentes's Terra Nostra", in: **Symposium; Washington** Vol. 57, N.º 2, (Summer), pp. 59-79

Casti, J., and Karlqvist, A., (eds.), (2003). *Art and complexity*. Elsevier

Corral, W. H., (2001). "Novelistas sin timón: exceso y subjetividad en el concepto de 'novela total', in: *MLN*, vol. 116, No. 2, March, pp. 315-349 (The Johns Hopkins University)

Cowan, G. A., Pines, D., Meltzer, D., (Eds.), (1999). *Complexity. Metaphors, Models, and Reality*. Cambridge, MA: Perseus Books

Forero Quintero, G., (2011). "La novela total o la novela fragmentaria en América Latina y los discursos de globalización y localización", in: *Acta Literaria*, No. 42, I sem., pp. 33-44

Garcés Valenzuela, J. (2010). "Escritores comprometidos, campo literario y novela total en los años sesenta. Mario Vargas Llosa, lector de Cien años de soledad", in: *Letras*; Lima, vol. 81, N.º 116, (Jan-Dec.), pp. 25-43

Hess, D. H., (1999). *Complexity in Maurice Blanchot's Fiction: Relations between science and literature.* Peter Lang Inc., International Academic Publishers

Medrano, M., E. H., (2012). Análisis y subversión del concepto de novela total en Los detectives salvajes de Roberto Bolaño", monography, department of literature,

Universidad Nacional, School of Human Sciences; available at: http://www.bdigital.unal.edu.co/9191/1/edgarhansmedranomora.2012.pdf

Omlor, D., (2014). "Mirroring Borges: The spaces of literature in Roberto Bolaño's 2666", in: *BHS*, 91.6, 659-670, doi:103828/bhs.2014.40

Ruiz-Pérez, I., (2016). "Elogio de la imaginación híbrida: Cervantes, Fuentes y la política de la novela total", in: *Romance Notes*, 56.1, pp. 53-59

Saenz, I., (1994). *Hacia la novela total: Fernando del Paso*. Madrid: Ed. Pliegos

Tiezzi. E., (2000). *La belleza e la scienza*. Raffaello Cortina

Zants, E., (1996). *Chaos theory, complexity, cinema, and the evolution of the French novel*. Edwin Mellen

Web pages:

*Le « Nouveau Roman » d'Annie Ernaux : un récit impossible ?*
http://www.fabula.org/lht/13/lavault.html

Vebret, J., (2012). "Pierre Jourde, Le Maréchal Absolu : le roman total", in: http://salon-litteraire.linternaute.com/fr/interviews/content/1799947-pierre-jourde-le-marechal-absolu-le-roman-total